# Multi-Particle Correlation among Grey particles Emitted in Nucleus-Nucleus Interactions

M.T. Hussein, N.M. Hassan, N.M. Sadek and Jamila Elsweedy
Physics Department, Faculty of Science, Cairo University, 12613 Giza, Egypt

## Abstract

The short range correlation among the emitted knock on nucleons from heavy ion collisions is used to reveal the dynamic characteristics of the reactions at medium and high energy collisions. Two and three particle correlations are considered in angular space to explain the emission of particles from collectively excited states of the nucleus as a Fermi liquid drop. Positive correlation is detected only among particles emitted in the extreme backward direction which is the coldest domain, and interpreted as direct non-statistical emission (splashing) of nucleons via the dynamical distortion of the Fermi surface accompanying the collective motion.

**Terminology**.
Nuclear Emulsion: Is used as a nuclear detector and target simultaneously. It displays traces during the passage of charged particle through it. It is considered as a complex target formed by two groups of nuclei, a light group (C,N,O) with average mass number equal 14, and a heavy one (AgBr) with average mass of 96 in addition to the hydrogen nuclei.
Grey particles: Those are the medium ionizing particles appear in grey color in emulsion detectors, with relative ionization density ~1.4 - 2. These particles are identified as knocked on protons.
Shower particles: Fast particles produced in emulsion detectors, mainly pions, having ionization density < 1.4.

## 1. Introduction

In previous works [1, 2] we demonstrate a model that describes the emission of grey particles in view of a thermodynamic picture. The assumption of local equilibrium as well as the concept of the canonical ensemble could explain the formation of a non-equilibrium thermodynamic nuclear matter in phase space. The formed nuclear system is divided into sub-domains each is characterized by a local characteristic temperature. The emitted grey particles are the superposition of the spectra emitted from the different domains. The question arising now passes around the dynamic effects acting on or playing a role in the process of the emission of the grey particles from the nuclear system. The default strategy deliberates with the correlation technique. The influence of short range correlations in the angular distribution of nucleons in nuclei can be evaluated assuming realistic dynamic effect. Many properties of nuclei could be understood within the independent particle model, where the nucleus is considered as a system of nucleons moving without residual interaction in a mean field or single particle potential. But the short range of the strong field and other tensor components that induce NN correlations in nuclear wave functions cannot be described by the independent particle model or even the Hartree-Fock approach. Various tools have been developed to account for these strong short-range correlations. These include variational calculations assuming Jastrow correlation functions [3], the correlated basis function method (CBF) [4], the exponential S method [5], the Brueckner-Hartree-Fock (BHF) approximation [6] and the self-consistent Green function approach [7]. In the present work we shall deal with the problem using the two-particle and the three-particle correlation functions, searching about a true signal in the angular space of the emitted nucleons that can express the presence of a dynamic effect.

## 2. Experimental Work

A stack of size 20 cm x 10 cm x 600 microns (undeveloped) was exposed to the beams of $^{24}$Mg nuclei at momentum of 4.5 A GeV/c. The stack was irradiated by beam oriented parallel to the length of the stack at the Dubna Synchro phasotron (Russia). The grain density, the blob density and the Lacunarity [8] of the emitted particles are measured and classified into shower (s), grey (g) and black (b) particles [9]. The charge and the mass of the emitted particles are identified by a couple of measurements (delta ray -energy measurements). In this work we are interested mainly in the g-particles. The energy of the g-particles [10] are measured by either the range or the Coulomb multiple scattering technique. A special computer program called (SRIM) [11, 12], is used to calculate the stopping and range of ions in energy (10eV – 2GeV/amu) in matter using a full quantum mechanical treatment of ion-atom (medium) collisions. The space angle of the identified singly charged grey particles were measured with accuracy less than $^{o}$0.1

## 3. The Short Range Correlations

A two-particle correlation function $R(z_1, z_2)$ is defined as:

$$R(z_1,z_2) = \left( \frac{1}{\sigma_{in}} \frac{d^2\sigma}{dz_1 dz_2} - \frac{1}{\sigma_{in}^2} \frac{d\sigma}{dz_1} \frac{d\sigma}{dz_2} \right) \bigg/ \left( \frac{1}{\sigma_{in}^2} \frac{d\sigma}{dz_1} \frac{d\sigma}{dz_2} \right) \quad (1)$$

$\sigma_{in}$, $\frac{d\sigma}{dz_1}$ and $\frac{d^2\sigma}{dz_1 dz_2}$ are the inelastic cross-section, the single- and the two-particle distributions, respectively, $z_1, z_2$ are the values of a certain physical quantity corresponding to the two particles. The correlation function measures the dependence of the production of the two particles on their own characteristic parameters. The positive high value of $R(z_1, z_2)$, means that the production of the particle "1" depends strongly on the production of particle "2", a zero value of $R(z_1, z_2)$ means that their production is independent. On the other hand a negative value of the correlation function means that the production of one of them prevents the production of the other.

The normalized inclusive correlation function can be written for the case $z = \cos\theta$ as

$$R(\cos\theta_1, \cos\theta_2)$$
$$= \left( \frac{1}{\sigma_{in}} \frac{d^2\sigma}{d\cos\theta_1 d\cos\theta_2} - \frac{1}{\sigma_{in}^2} \frac{d\sigma}{d\cos\theta_1} \frac{d\sigma}{d\cos\theta_2} \right) \bigg/ \left( \frac{1}{\sigma_{in}^2} \frac{d\sigma}{d\cos\theta_1} \frac{d\sigma}{d\cos\theta_2} \right)$$
$$= \frac{NN_2(\cos\theta_1, \cos\theta_2)}{N_1(\cos\theta_1) N_1(\cos\theta_2)} - 1$$

$$(2)$$

$N_1(\cos\theta_1)$ is the number of interactions having a grey particle between $\cos\theta_1$ and $\cos\theta_1 + d\cos\theta_1$, $N_2(\cos\theta_1, \cos\theta_2)$ is the number of interactions having at least two grey particles the first with angle between $\cos\theta_1$, $\cos\theta_1 + d\cos\theta_1$, and the other of angle between $\cos\theta_2$, $\cos\theta_2 + d\cos\theta_2$ and $N$ is the total number of investigated inelastic interactions. By analogy, the three particle correlation function is defined as:

$$R(\cos\theta_1, \cos\theta_2, \cos\theta_3)$$

$$= \left( \begin{array}{c} \frac{1}{N} N_3(\cos\theta_1, \cos\theta_2, \cos\theta_3) + 2\frac{1}{N^3} N_1(\cos\theta_1) N_1(\cos\theta_2) N_1(\cos\theta_3) \\ -\frac{1}{N^2} N_2(\cos\theta_1, \cos\theta_2) N_1(\cos\theta_3) - \frac{1}{N^2} N_2(\cos\theta_2, \cos\theta_3) N_1(\cos\theta_1) \\ -\frac{1}{N^2} N_2(\cos\theta_3, \cos\theta_1) N_1(\cos\theta_2) \end{array} \right) *$$

$$\left( \frac{1}{N^3} N_1(\cos\theta_1) N_1(\cos\theta_2) N_1(\cos\theta_3) \right)^{-1}$$

(3)

Figure (1) shows the two particle correlation function for grey particles emitted in Mg-Em interactions at 4.2 A GeV/c. A clear signal of positive correlation is observed only for the enormously backward emitted particles. In fact this signal has no substantial value, since the cross section of producing grey particles at this angle is very weak as appears in Figure (2). Nevertheless, this backward signal may be interpreted as the so called side splash effect; the region of the occurrence depends on the temperature of the domain where the grey particles are produced. The previous results of the thermodynamic model [2, 13, 14] showed that the angular distribution of the grey particles is strongly correlated to their emission energy, so that fast energetic particles are emitted in the forward direction from domains characterized with high temperature. Accordingly, it is reasonable to interpret Figure (1) as the increase of the value of the two particle correlation function toward the colder domain, and a true signal is recorded at the coldest zone corresponding to the extreme backward emission. The emission of particles from a collectively excited state of the nucleus as a Fermi liquid drop can occur in two ways. First, due to the relaxation processes where the collective energy is transferred to the intrinsic degrees of freedom with subsequent evaporation of particles. On the other hand, a direct non-statistical emission (splashing) of nucleons is also possible via the dynamical distortion of the Fermi surface accompanying the collective motion. In general the relative contributions of these mechanisms depend upon the magnitude of the nuclear friction coefficient. The limiting cases of the direct (non-statistical) particle emission from the non-damped giant multipole resonance (GMR) and the particle evaporation from the heated nucleus may also be considered. The direct particle emission from the GMR has been extensively studied within pure quantum mechanical approaches [15, 16].

Particles emitted from excited nucleus due to both the evaporation and the splashing (emission from a cold vibrating nucleus) are due to the collective motion of the nuclear Fermi liquid and are accompanied by direct non-statistical emission of nucleons via the dynamical distortion of the Fermi surface so they are responsible to the presence of the short range correlation.

The results displayed in Figure (3) show that the three particle correlation function has strong negative correlation in the backward direction. This means that the short range correlation is observed only among the two particles and not among the three grey particles. This means that the splashes come out from the cold nuclear medium are emitted in fine form.

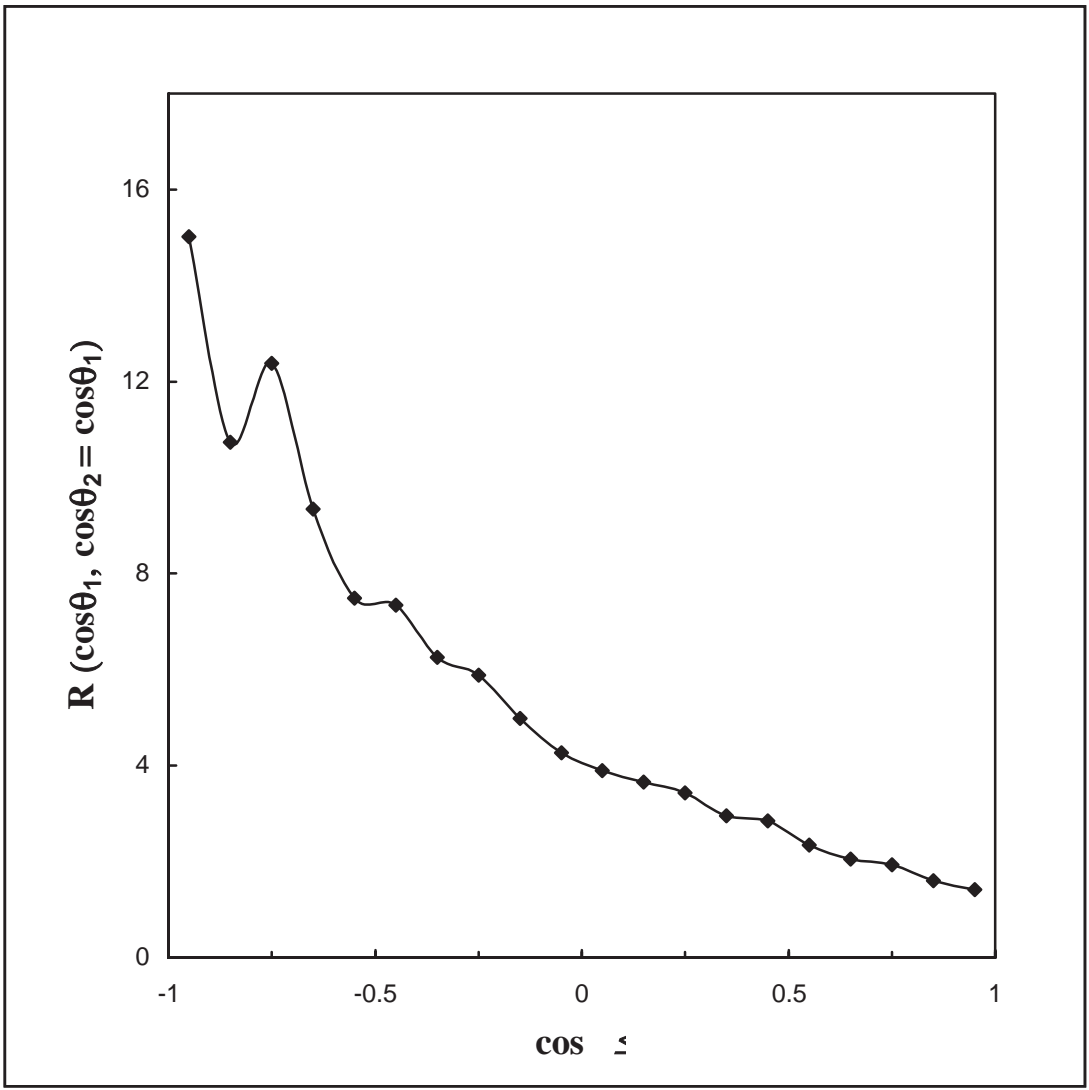

Fig. (1)

Two particle correlation function for g-particles emitted in Mg-Em interactions at 4.2 A GeV projectile energy.

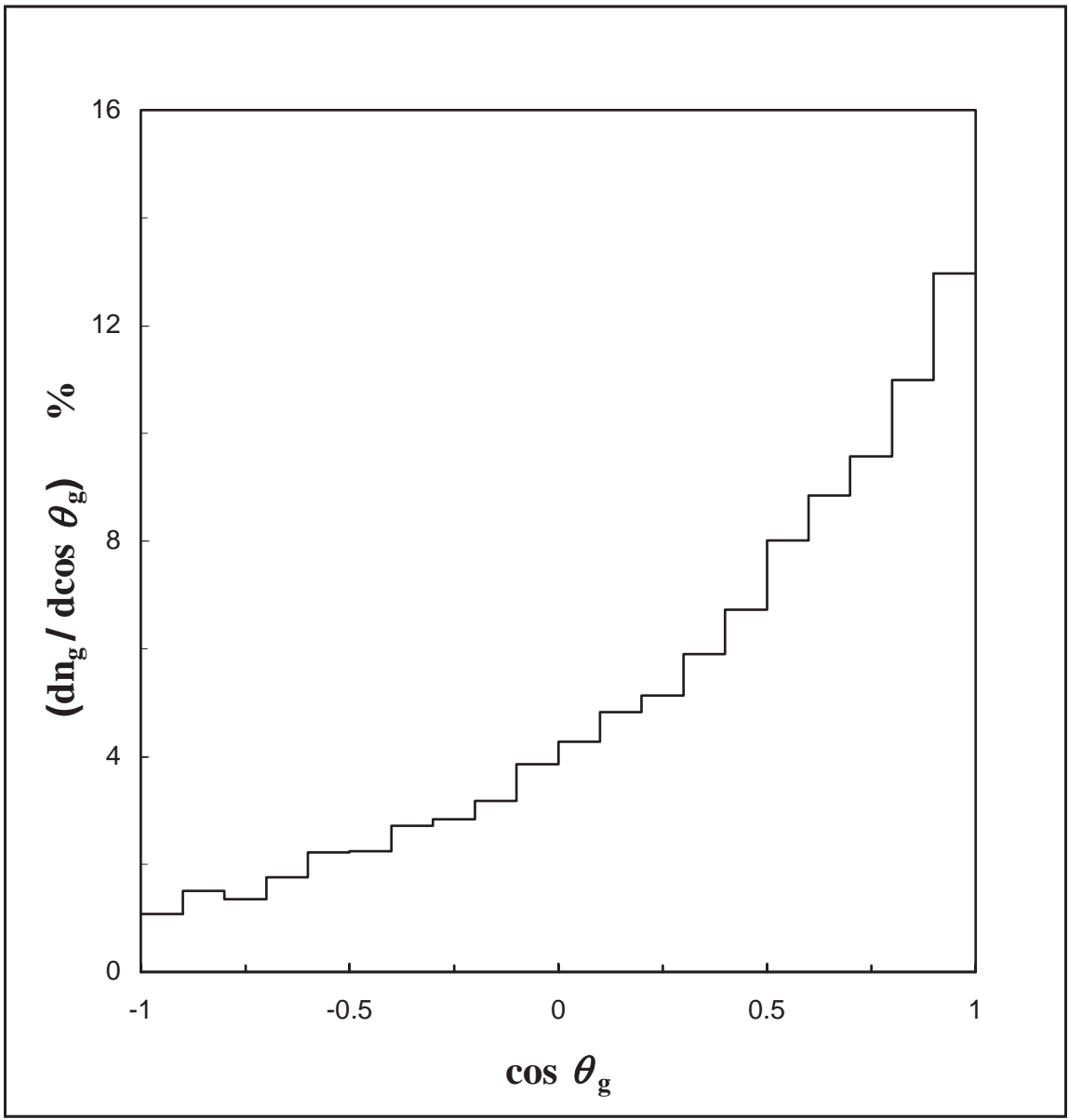

**Fig. (2)**

Angular distribution function for g-particles emitted in Mg-Em interactions at 4.2 A GeV projectile energy.

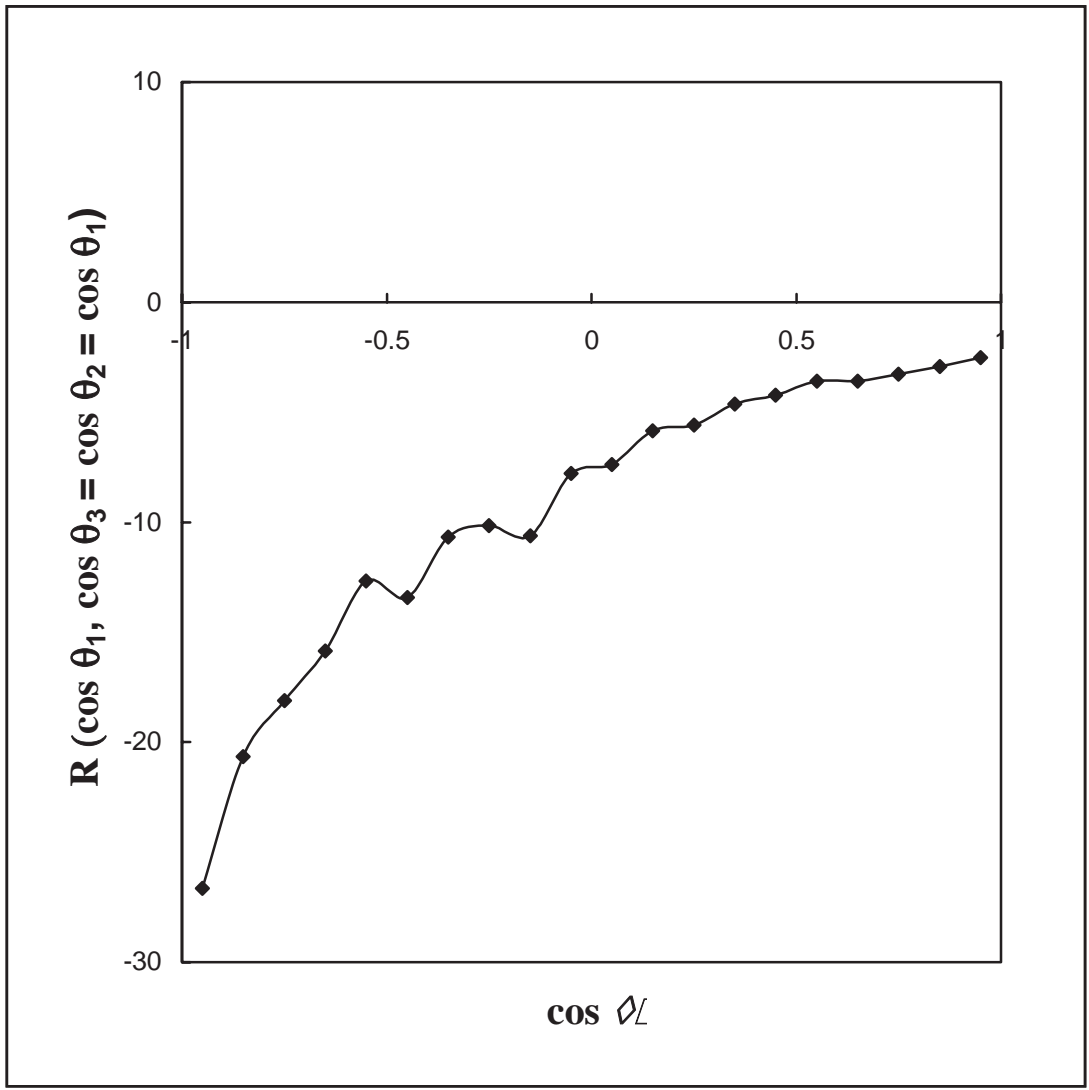

**Fig. (3)**

Three particle correlation function for g-particles emitted in Mg-Em interactions at 4.2 A GeV projectile energy.